\documentclass[11pt]{article}
\pdfoutput=1
\usepackage{jheppub}
\usepackage{graphicx}
\usepackage{color, bm, enumitem}

\newcommand{\Tr}{{\text{Tr}}}

\newcommand{\be}{\begin{equation}}
\newcommand{\ee}{\end{equation}}
\newcommand{\bp}{\begin{pmatrix}}
\newcommand{\ep}{\end{pmatrix}}
\newcommand{\bsp}{\left(\begin{smallmatrix}}
\newcommand{\esp}{\end{smallmatrix}\right)}

\newcommand{\C}{{\mathbb C}}

\newcommand{\CN}{{\mathcal N}}

\title{Curious Aspects of Three-Dimensional ${\cal N}=1$ SCFTs}
\author[1]{Davide Gaiotto,}
\author[2,3]{Zohar Komargodski,}
\author[1]{and Jingxiang Wu}

\affiliation[1]{Perimeter Institute for Theoretical Physics, 31 Caroline St.\,N., Waterloo, ON N2L 2Y5, Canada}
\affiliation[2]{Department of Particle Physics and Astrophysics, Weizmann Institute of Science, Israel}
\affiliation[3] {Simons Center for Geometry and Physics, Stony Brook University, Stony Brook, NY}

\abstract{  We study the dynamics of certain 3d ${\cal N}=1$ time reversal invariant theories.
Such theories often have exact moduli spaces of supersymmetric vacua. We propose several dualities and we test these proposals by comparing the deformations and supersymmetric ground states. First, we consider a theory where time reversal symmetry is only emergent in the infrared and there exists (nonetheless) an exact moduli space of vacua. This theory has a dual description with manifest time reversal symmetry. Second, we consider some surprising facts about ${\cal N}=2$ $U(1)$ gauge theory coupled to two chiral superfields of charge 1. This theory is claimed to have emergent $SU(3)$ global symmetry in the infrared. We propose a dual Wess-Zumino  description (i.e. a theory of scalars and fermions but no gauge fields) with manifest $SU(3)$ symmetry but only ${\cal N}=1$ supersymmetry. We argue that this Wess-Zumino model must have enhanced supersymmetry in the infrared. Finally, we make some brief comments about the dynamics of ${\cal N}=1$ $SU(N)$ gauge theory coupled to $N_f$ quarks in a time reversal invariant fashion. We argue that for $N_f<N$ there is a moduli space of vacua to all orders in perturbation theory but it is non-perturbatively lifted.
}

\begin{document}
\today
\maketitle

\section{Introduction} 
In theories with four supercharges (both in 2+1 and 3+1 dimensions) it is well known that there are powerful non-renormalization theorems~\cite{Grisaru:1979wc,Seiberg:1993vc,Seiberg:1994bz}. This is due to the holomorphic nature of the superpotential. These non-renormalization theorems have many applications. In particular, they allow in many cases to determine the space of supersymmetric ground states. In such theories with four supercharges, these spaces are complex (in fact, K\"ahler) manifolds. Understanding the space of ground states is a crucial step before one can study the behavior near various interesting singular points.

Here we will study 2+1 dimensional theories with $\mathcal{N}=1$ supersymmetry, namely, two real supercharges. These theories have a real superpotential, $W$. Since one does not have complex analysis at one's disposal, it is typically hard to find exact results. It would seem that $W$ can 
receive corrections since it is easy  to write real functions invariant under the global symmeties. One aspect of this problem was recently studied in~\cite{Bashmakov:2018wts}, where it was shown that $\mathcal{N}=1$ theories often have walls in parameter space and it is possible to obtain exact results near these walls using only the leading radiative correction to $W$. The models studied~\cite{Bashmakov:2018wts} do not have time reversal symmetry due to various Chern-Simons terms. Some related aspects of 
Chern-Simons-matter theories with ${\cal N}=1$ supersymmetry were recently discussed in~\cite{Benini:2018umh}. 

Here we will study some $\mathcal{N}=1$ models with time reversal symmetry. Such models often have exact real continuous manifolds (with singularities) of supersymmetric ground states. A related fact is that the renormalization of $W$ is severely restricted, and often $W$ cannot be corrected at all. These non-renormalization theorems 
are due to the fact that the superpotential is a pseudo-scalar under time reversal symmetry. This property of the superpotential was already noted in~\cite{Affleck:1982as}. Here we will review this fact and develop some applications of it. 

As a simple example, consider the model of 3 real scalars and three Majorana fermions, embedded into the three real superfields $A,B,C$ with superpotential 
$$W=ABC~.$$
We argue that the full theory has a moduli space of vacua consisting of 3 real lines that intersect at an $\CN=1$ SuperConformal Field Theory (SCFT). 

Next, we will study the theory of a charge 2 superfield coupled to a $U(1)$ gauge field in a time reversal invariant fashion. This theory has $\mathcal{N}=1$ and $\mathcal{N}=2$ versions and we find a dual description in both cases. The dual description consists of a pure $U(1)_2$ TQFT tensored with  
a charge 1 superfield coupled to a $U(1)$ gauge field with a Chern-Simons term at level $3/2$. Therefore, loosely speaking (the details will be presented in the main body of the paper), the duality is 
$$U(1)_0+{\rm charge \ 2} \longleftrightarrow U(1)_2\otimes\left[U(1)_{3/2}+{\rm charge\ 1}\right]$$
Time reversal symmetry in the dual description on the right hand side is therefore emergent in the infrared. 
Interestingly, the  theory on the right hand side 
has a moduli space of $\mathcal{N}=1$ vacua even though it has no microscopic time reversal symmetry.  We will explain the basic mechanism that allows such exact moduli spaces of vacua to exist without time reversal symmetry.

We then discuss a new duality between an $\mathcal{N}=2$ SQED theory and a Wess-Zumino model. This duality has two surprising aspects. First, the symmetry of the infrared fixed point is enhanced from $U(2)$ to $SU(3)$. Second, the dual theory is a 
Wess-Zumino like model but we do not have an $\mathcal N=2$ description of it. The $\mathcal{N}=2$ supersymmetry arises in the infrared and only an $\mathcal{N}=1$ symmetry is manifest in the flow 
$$\mathcal{N}=2 \ \ \ U(1)+2\ {\rm charge} \ 1 \longleftrightarrow  \mathcal{N}=1 \ \ \ W= \Tr \Phi^3~,$$
where $\Phi$ is in the adjoint of $SU(3)$ (i.e. 8 real scalar degrees of freedom). $W$ is an $\mathcal{N}=1$ superpotential. We see that on the right hand side there is emergent supersymmetry (and $R$-symmetry) in the infrared and on the left hand side there is emergent $SU(3)$.

This duality also has a purely $\mathcal{N}=1$ version (where the Wess-Zumino model has 7 real scalar fields rather than 8), but in that case there is no enhanced global symmetry and no enhanced supersymmetry. However,  there is a moduli space of vacua, which we match across the duality.

We close with brief remarks on ${\cal N}=1$ time reversal invariant non-Abelian gauge theories. More specifically, we consider $SU(N)$ gauge theories minimally coupled to $N_f$ fundamental multiplets. We show that for $N_f<N$ there is a moduli space of vacua to all orders in perturbation theory but it is non-perturbatively lifted.

The outline of this note is as follows. In section 2 we explain how time reversal symmetry acts in the context of $\mathcal{N}=1$ supersymmetry. We give some examples of applications of the fact that $W$ is a pseudo-scalar. 
 In section 3 we discuss the theory of a charge 2 particle coupled to a $U(1)$ gauge field. We discuss the $\mathcal{N}=1$ and $\mathcal{N}=2$ versions of the theory and find dual descriptions in both cases. We outline the connection of these dualities to some non-supersymmetric dualities. 
 In section 4 we discuss QED with two charge 1 particles, and again discuss $\mathcal{N}=1$ and $\mathcal{N}=2$ versions of the theory, finding dualities in both cases, and in particular, in the latter case, we find an enhanced global symmetry in the infrared. On the other side of the duality, we find enhanced supersymmetry (and $R$ symmetry) in the infrared. 
In section 5 we make some comments about $\mathcal{N}=1$ supersymmetric non-Abelian $SU(N)$ gauge theories with time reversal symmetry. We show that for $N_f<N$ there is a moduli space of vacua to all orders in perturbation theory 
but supersymmetry is  broken non-perturbatively on this moduli space.
  
\section{The Action of Time Reversal Symmetry}
We take the sigma matrices to be as usual 
\begin{equation} \sigma^1=\left(\begin{matrix}0 & 1 \\ 1 & 0 \end{matrix}\right)~,\quad \sigma^2=\left(\begin{matrix}0 & -i \\ i & 0\end{matrix}\right) ~,\quad \sigma^3=\left(\begin{matrix}1 & 0 \\ 0 & -1 \end{matrix}\right)~.  \end{equation} 
Below we will use the Lorentzian signature $(-,+,+)$. We will denote the corresponding $\gamma$ matrices by $\gamma^{0,1,2}$
\begin{equation} \gamma^0=i\sigma^2~,\qquad \gamma^1=\sigma^1 ~,\quad \gamma^2=\sigma^3~. \end{equation} 

A Majorana spinor is a real two dimensional vector $\lambda_\alpha$. 
We define $\bar \lambda\equiv \lambda^T\gamma^0$. As usual, $\bar \lambda\lambda$ is a Lorentz invariant and $\bar \lambda\gamma^\mu \partial_\mu \lambda$ is likewise a Lorentz invariant. In our conventions, when we add these terms to the action, they both have to be multiplied by a factor of $i$. 

Time reversal symmetry acts as follows (in addition to the obvious action on space-time coordinates, where the sign of $x^0$ is reversed): 
\begin{equation}\label{Taction} T:  \lambda\to \pm\gamma^0\lambda~.\end{equation}
One is free to choose the sign in this transformation rule. It is important to remember that $T$ is anti-linear. This will be used throughout below.  
 The Majorana mass term is odd under time reversal symmetry (whatever sign in~\eqref{Taction} we use)
$T(i\bar \lambda \lambda)=i\lambda^T\gamma^0\gamma^0\gamma^0\lambda =-i\bar \lambda\lambda$. The kinetic term is of course even under time reversal symmetry.\footnote{ 
The action on the kinetic term is $T(i\bar \lambda\gamma^\mu\partial_\mu \lambda)=T(i\bar \lambda\gamma^0\partial_0 \lambda)+T(i\bar \lambda\gamma^i\partial_i \lambda)=
(-i)(-1)^2\lambda^T\gamma^0\gamma^0\gamma^0\partial_0 \gamma^0\lambda+(-i)(-1) \lambda^T\gamma^0\gamma^0\gamma^i\partial_i\gamma^0 \lambda=i\bar\lambda\gamma^0\partial_0\lambda+i \bar\lambda\gamma^i\partial_i \lambda=i\bar\lambda\gamma^\mu\partial_\mu\lambda$. Therefore, the kinetic term is time reversal invariant.}

In theories with $\mathcal{N}=1$ supersymmetry, the superspace consists of the usual coordinates $x^\mu$ and the Majorana Grassmann coordinates $\theta_\alpha$. Since the $\theta_\alpha$ are Majorana, time reversal symmetry must act on them as in~\eqref{Taction}: $T: \theta\to \pm \gamma^0\theta$.
The Lagrangian is determined by the real superpotential $W$, 
$$\mathcal{L}=i \int d^2\theta \ W~.$$
The factor of $i$ is important for the theory to have a Hermitian Hamiltonian.
As with the fermion mass,~\eqref{Taction} implies that $id^2\theta$ is odd under time reversal symmetry. Therefore to write time reversal invariant theories we need $W$ to be a {\it pseudo-scalar} 
\begin{equation}\label{pseudos}T: W\to -W~.\end{equation}
In addition, $W$ clearly needs to be invariant under all the global symmetries and gauge symmetries. 
These conditions will turn out to be very restrictive, as we will see.

Let $R$ be a real superfield (therefore it contains a real boson and a Majorana fermion). The bottom component of $R$ will be denoted by $R\bigr|$. Suppose that $R$ is a scalar, that is, under time reversal symmetry $T: R \to R$. Clearly we cannot write any time reversal invariant superpotential $W(R)$ which does not contain superspace derivatives\footnote{Below when we establish various non-renormalization theorems, except when mentioned explicitly, we are always concerned with the terms in the superpotential which do not vanish for constant bottom components, i.e. potential terms. Of course, also the terms with superspace derivatives need to obey the same selection rules under time reversal symmetry.} since there is no way to make it odd under time reversal symmetry as required in~\eqref{pseudos}.
Of course, without supersymmetry we could easily write such a potential, but the fermionic terms accompanying any nontrivial function $W(R)$ are necessarily odd and hence we cannot write a potential consistent with $\mathcal{N}=1$ SUSY
\begin{equation}\label{nosup}T: R \to R~,\qquad W(R)=0~.\end{equation}
In particular, any time reversal invariant theory of a real scalar superfield must have an exact real flat direction.

If $R\bigr|$ is a pseudo-scalar, $T: R\to -R$, then we can easily write analytic superpotentials (without superspace derivatives) such as $W(R)=R+R^3+...$, containing only odd powers of $R$ and thus preserving time reversal symmetry  
\begin{equation}\label{oddsup}T: R \to -R~,\qquad W(R)=R+R^3+\cdots~.\end{equation}

The most general case can always be analyzed by simply decomposing the superfields into real superfields and using the rules above.

If we start with a theory with $\mathcal{N}=2$ supersymmetry in 2+1 dimensions, we can rewrite it as an $\mathcal{N}=1$ theory. The $\mathcal{N}=1$ superpotential would be a pseudo-scalar but the original $\mathcal{N}=2$ superpotential may or may not be a pseudo-scalar, depending on how the time reversal symmetry is defined to act. 

\subsection{The $ABC$ Model} 
Let us consider a toy model which exhibits some of the ideas above. Take three real superfields $A,B,C$ and consider the superpotential 

\begin{equation}\label{ABCmodel}W= gABC~.\end{equation} 
($g$ is the coupling constant). This is an interacting, super-renormalizable model; it is strongly coupled in the infrared.

First let us consider the classical supersymmetric ground states. The equations for those are 
$$AB=AC=BC=0~,$$
and hence there are solutions with $A=B=0$ and arbitrary (real) $C$, $A=C=0$ and arbitrary (real) $B$, $B=C=0$ and arbitrary (real) $A$. These are three real lines which intersect at the origin. Let us consider the fate of this moduli space quantum mechanically. 

The global (unitary) symmetries of the model are generated by the $\mathbb{Z}_2$ action
$A\to -A,B\to -B, C\to C$ and by the permutation symmetry $S_3$ acting on $A,B,C$. In total, this group has 24 elements. This group turns out to be isomorphic to $S_4$.\footnote{We thank M.~Rocek for making this observation.}
This is not yet sufficiently constraining as this by itself would allow as we integrate out high momentum mode (near the ultraviolet, where the original degrees of freedom are still useful) to generate new vertices such as $A^4+B^4+C^4$, which is indeed invariant under the $S_4$. Such a term would lift the moduli space. 

Now let us add time reversal symmetry into our considerations. Since the superpotential has to be odd under time reversal symmetry, we can by no loss of generality choose $A$ to be a pseudo-scalar superfield and $B,C$ are scalar superfields. All the other choices can be obtained by composing this time reversal symmetry with $S_4$.

This significantly restricts the vertices that can be generated as we integrate out high energy mode in the ultraviolet (where, again, the original degrees of freedom are still useful). For example,  $A^4+B^4+C^4$ cannot be generated since it is even rather than odd under time reversal symmetry. Similarly, no vertex of the sort $A^{n}+B^{n}+C^{n}$ with integer $n$ can be generated. We may, however, generate various new irrelevant operators such as  
$$\sim \int d^2\theta\ ABC(A^2+B^2+C^2)~,$$
which is both irrelevant and does not lift the moduli space. It is indeed easy to see that every vertex that is generated as we perform the renormalization group transformations has to be proportional to $ABC$ and hence cannot lift the moduli space.\footnote{To see that it is convenient to think about the generated vertices as some polynomials in $A,B,C$. If any of the terms does not contain all the three fields, we can choose the time reversal symmetry to be such that it acts on all the fields in that vertex as scalars and hence such a vertex must violate some combination of the time reversal symmetry and the unitary symmetries. We thank M. Rocek for a discussion of this point.} 
 
 It is  easy to prove that the classical moduli space continues to exist non-perturbatively by going on the moduli space and computing the Coleman-Weinberg~\cite{Coleman:1973jx} potential. Let us consider the branch, where, without loss of generality $B=0,C=0$, and $A$ is arbitrary. Then, we can integrate out the heavy $B,C$ fields and we obtain some effective superpotential $W_{eff}=W_{eff}(A)$. We can choose $A$ to be a scalar under time reversal symmetry and hence $W_{eff}=0$ necessarily follows. (Time reversal symmetry is not spontaneously broken for large enough $A$ since the theory is arbitrarily weakly coupled there.\footnote{We thank N.~Seiberg for a discussion of this point.})

The moduli space of 3 real lines meeting at a point therefore survives in the full quantum theory. At the intersection there is an $\mathcal{N}=1$ SCFT. The model has no relevant deformations which preserve all the symmetries and $\CN=1$ supersymmetry. 
We can deform the model by the mass term $\int d^2\theta\ m\left(A^2+B^2+C^2\right)$ such that $m$ is odd under time reversal but all the global symmetries are maintained. For either positive or negative $m$ we have 5 gapped trivial vacua. These 5 gapped vacua split into one that preserves $S_4$ and in the other 4 vacua the unbroken symmetry is isomorphic to $S_3$ and therefore these 4 vacua are related by the symmetry breaking  pattern
$$S_4\longrightarrow S_3~.$$ 

\section{$\mathcal{N}=1$ Abelian Gauge Theory with a Charge 2 (Super)Field}

Here we consider a $U(1)$ gauge field (with no Chern-Simons term) coupled minimally ($\mathcal{N}=1$ supersymmetrically) to a charge 2 multiplet $\Phi$.\footnote{A single charge 1 fermion cannot be coupled to a dynamical $U(1)$ gauge field while preserving time reversal symmetry because of an (ABJ-like) anomaly~\cite{Niemi:1983rq,Redlich:1983kn,Redlich:1983dv}. This is why we study the model of a $U(1)$ gauge field coupled to a charge 2 multiplet, which suffers from no such anomaly and time reversal symmetry can be maintained. The non-supersymmetric version of this model was considered in~\cite{Cordova:2017kue} and the non-supersymmetric monopole-deformed version in~\cite{Gomis:2017ixy}.} The theory is time reversal invariant, which means that the Chern-Simons level vanishes (hence the subscript 0 on the gauge group).
We are using the usual convention, where integrating out a charge 1 fermion shifts the Chern-Simons level by $\pm\frac12$ depending on the sign of the mass.  

The most general superpotential is again some function $W=W(|\Phi|^2)$. Hence, there is no way to write a superpotential that preserves time reversal symmetry. This leads to a non-renormalization theorem:
If we start from a superpotential that vanishes at tree level, then the superpotential vanishes in the full quantum theory as we integrate out high energy degrees of freedom near the ultraviolet. We see that in this case the implication of time reversal symmetry is even stronger than in the $ABC$ model.

The exact vanishing of the superpotential shows that the time reversal invariant theory has a moduli space given by arbitrary expectation values of $\Phi\bigr|$ divided by the gauge symmetry. This leads to a moduli space isomorphic to $\mathbb{R}_+$
\begin{equation}\label{msRplus}\mathcal{M}_{vac}\simeq \mathbb{R}_+ ~. \end{equation}
The effective theory on this $\mathbb{R}_+$ requires some attention. At the origin of $\mathbb{R}_+$ there is a certain SCFT. Let us now consider what happens away from the origin.
Due to the fact that $\Phi$ has charge 2 under the gauge symmetry, there is an unbroken $\mathbb{Z}_2$ gauge theory on the moduli space. The most familiar version of $\mathbb{Z}_2$ gauge theory is described by the ${\bf k}$ matrix 
$$\bf{k}=\left(\begin{matrix} 0 & 2 \\ 2 & 0 \end{matrix}\right)~,$$
however, here we have a Dijkgraaf-Witten modification~\cite{Dijkgraaf:1989pz} of this $\mathbb{Z}_2$ gauge theory. The easiest way to understand it is through the fact that our theory has a vanishing quantum Chern-Simons level (which is why the theory is time reversal invariant), but the bare Chern-Simons level (in absolute value) is $2$.

This modification of $\mathbb{Z}_2$ gauge theory can be described by the ${\bf k}$ matrix 
\begin{equation}\label{DWmodified}\bf{k}=\left(\begin{matrix} 2 & 2 \\ 2 & 0 \end{matrix}\right)~.\end{equation}
It is easy to see that this modified $\mathbb{Z}_2$ gauge theory is isomorphic to $U(1)_2\times U(1)_2$.\footnote{Here and below we will use the fact that $U(1)_2\simeq U(1)_{-2}$ as spin TQFTs.}
 Therefore we conclude that on the moduli space~\eqref{msRplus} the effective theory consists classically of the modulus $\rho$ parameterizing $\mathbb{R}_+$ as well as the TQFT $U(1)_2\times U(1)_2$. The origin is a singular point, where there is a SCFT. The TQFT on the moduli space is time reversal invariant (see footnote~7), which is of course important for the consistency of our picture.

Let us now break time reversal symmetry explicitly by adding a mass term in the superpotential 
 $$W=m|\Phi|^2~.$$
For $m>0$ it flows in the infrared to $U(1)_2$ TQFT and of course also for $m<0$ it flows to the $U(1)_2$ TQFT. It is important to remark that the Witten index at negative $m$ is $-2$ while at positive $m$ it is $+2$ and therefore the Witten index jumps. It jumps at $m=0$ by the appearance of the exact moduli space. (Hence, the index at strictly $m=0$ is ill defined.)

It will prove useful to make a few comments about the model where we also add a quartic term
\begin{equation}\label{quartic}W=m|\Phi|^2+\lambda|\Phi|^4~.\end{equation}
For fixed nonzero $\lambda$ there is now no moduli space at $m=0$. If $\lambda>0$ we have at positive $m$ one vacuum with $U(1)_2$ TQFT and for negative $m$ we have two ground states, one with $U(1)_2$ TQFT and one with $U(1)_2\times U(1)_2$ TQFT. For $\lambda<0$ the situation is simply reversed.
 
 These various phases as well as the moduli space of vacua (at $\lambda=0$) with the particular TQFT on the moduli space~\eqref{DWmodified} will be later compared to a dual description of this theory.
 
To understand the duality in the next subsection it is useful to extend the theory~\eqref{quartic} even further. 
We add another neutral real scalar superfield $S$ and consider the theory as a function of some superpotential terms for the superfield $S$. 
The tree-level superpotential is given by 
$$W=S|\Phi|^2-\frac12\tilde M S^2+\tilde\xi S$$
and we study the model as a function of $\tilde\xi$ and $\tilde M$. Time reversal symmetry forces $S$ to be a pseudo-scalar (this we see from the first term in the superpotential) and as a result the parameter $\tilde M$ is time reversal odd and the parameter $\tilde\xi$
is time reversal even. Quantum corrections to the superpotential are allowed. We will keep only the one-loop correction for $S$ and later explain why this is sufficient for our purposes.  The one-loop corrected superpotential is given by 
 \begin{equation}\label{oneloop} W_{{\rm Tree+1-Loop}}=S|\Phi|^2-\frac12\tilde M S^2-\frac12S|S|+\tilde\xi S~.\end{equation}
 In the limit of large $|\tilde M|$ the model clearly reduces to the minimal model we analyzed before, with a $|\Phi|^4$ correction in the superpotential~\eqref{quartic}. When we take $|\tilde M|$ to be strictly infinite then we recover precisely the minimal model of a charge 2 multiplet with a vanishing superpotential and moduli space of vacua isomorphic to $\mathbb{R}_+$.
 
A very important fact to note is that the model has enhanced $\CN=2$ supersymmetry (and $R$ symmetry) for $\tilde M=0$.
The phases of the model as a function of $\tilde M$ and $\tilde \xi$ are 

\begin{itemize}

\item $\tilde M>1$. For positive $\tilde\xi$ we have a gapped SUSY vacuum with $U(1)_2$ TQFT in the deep infrared.  For negative $\tilde\xi$ we have two gapped SUSY vacua, one with the $\mathbb{Z}_2$ gauge theory (recall that because of the Dijkgraaf-Witten term it is isomorphic to $U(1)_2\times U(1)_2$) gauge theory and one with $U(1)_2$ TQFT. The Witten index is constant as a function of $\xi$. 
\item $\tilde M<-1$. For positive $\tilde \xi$ we have a SUSY vacuum with $U(1)_2$ TQFT.  For negative $\tilde\xi$ we have two SUSY vacua, one with $U(1)_2$ TQFT and one with $U(1)_2\times U(1)_2$ TQFT.  The Witten index is again constant as a function of $\xi$.
These are the same phases that we saw for $\tilde M>1$. This is not surprising, since under time reversal symmetry 
$\tilde M\to -\tilde M$ (and $\tilde \xi$ is fixed).
\item $|\tilde M|<1$. For positive $\tilde\xi$ we have two  gapped SUSY vacua, each carrying a $U(1)_2$ TQFT. For negative $\tilde\xi$ we have a gapped SUSY vacuum with $U(1)_2\times U(1)_2$ TQFT. The Witten index is constant as a function of $\tilde\xi$.

\end{itemize}

We draw all these phases in the following figure. 

\begin{center}\includegraphics[scale=0.3]{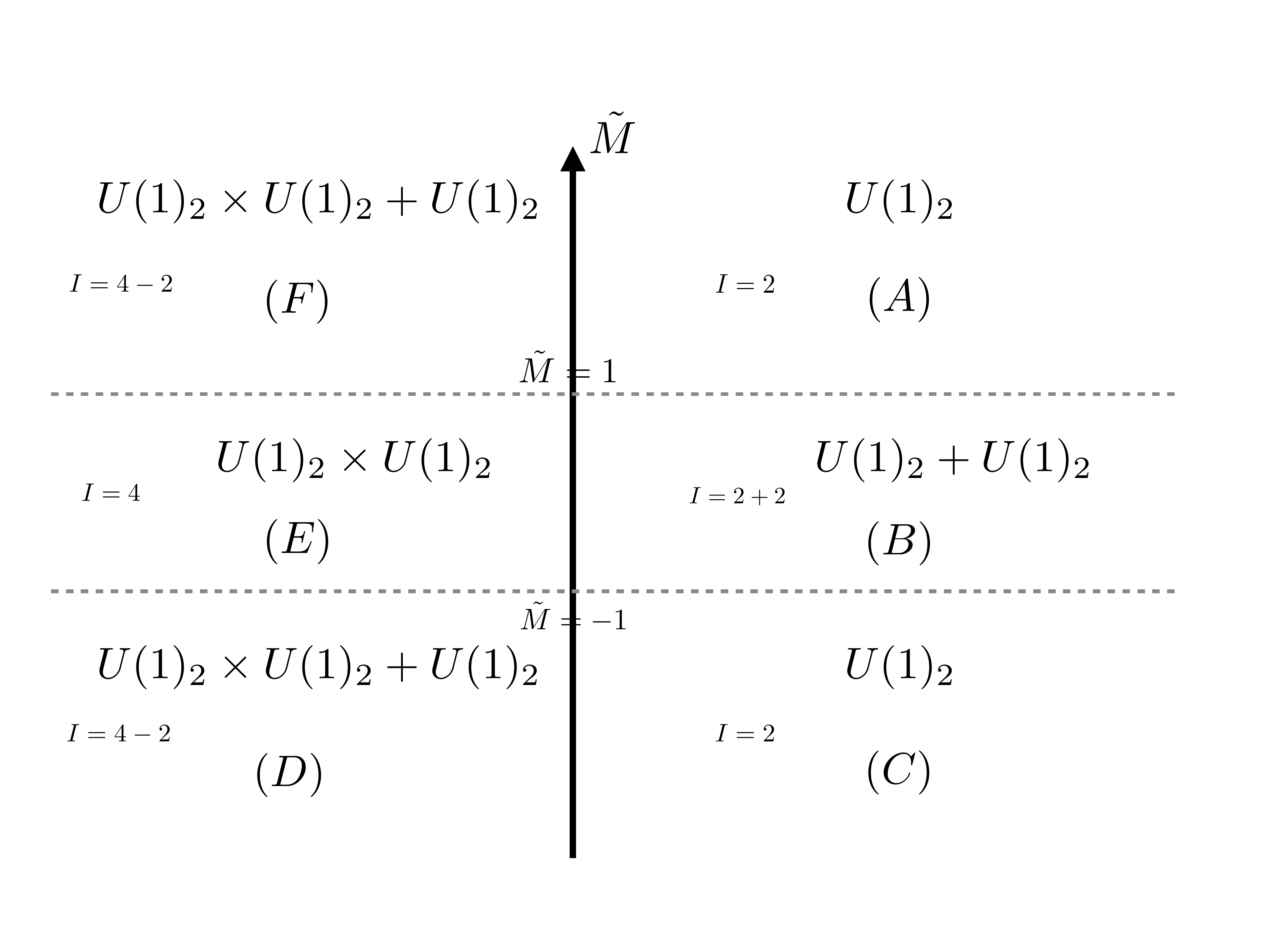}\end{center}

There are 6 regions, labeled A,B,C,D,E,F clockwise and in each region we record the vacua and their Witten indices. The bold vertical line is at $\tilde\xi=0$, positive $\tilde\xi$ is to the right and negative $\tilde\xi$ is to the left. 

While as a function of $\tilde\xi$ the Witten index does not jump, if we travel vertically on the diagram (i.e. change $\tilde M$) the Witten index jumps and there are two walls at $\tilde M=\pm 1$. The walls may move as we include higher loop corrections, but the number of such regions and the (gapped) vacua in each region would not change, which is why we are allowed to keep just the one loop correction in the superpotential. An important fact  is that $\tilde M=\infty$ and $\tilde M=-\infty$ should be really identified (with a twist in the horizontal direction), as they describe the theory~\eqref{quartic} with $\lambda=0$. We also see that at large $|\tilde M|$ we find precisely the phases we have seen in~\eqref{quartic} for positive and negative $\lambda$.

\subsection{A Dual Description with Emergent Time Reversal Symmetry}

We begin with a quick review of the $\mathcal{N}=1$ supersymmetric  model of a charge 1 superfield $\tilde \Phi$ coupled minimally to a $U(1)_{3/2}$ gauge field. 
We study the phases of the model as a function of the mass in the superpotential $W=m|\tilde\Phi|^2$. In~\cite{Bashmakov:2018wts} 
it was shown that there is a wall at $m=0$ (i.e. the Witten index jumps there due to the appearance of a new vacuum at infinity) and due to a radiative correction there is a conformal field theory at some $m_*>0$ where 
two trivial vacua merge into a supersymmetric vacuum with $U(1)_2$ TQFT.

Here we will generalize this model a little bit and add an additional neutral field $S$ as well as a superpotential
 $W=S|\tilde\Phi|^2$. This model naturally has two deformation parameters, $M$ and $\xi$
 $$W_{{\rm Tree}}=gS|\tilde \Phi|^2+\frac12MS^2+\xi S~.$$
  This model has no microscopic time reversal symmetry. 
If we take $|M|$ to be very large we can integrate out $S$ and the model reduces to the minimal model without $S$ and some quartic coupling $|\tilde \Phi|^4$.

To study the critical points of this superpotential, it is important to add the leading nontrivial radiative correction for $S$: 
\begin{equation}\label{olc}W_{One-Loop}=gS|\tilde\Phi|^2-\frac12Mg^2S^2+\xi S-\frac14g^2S|S|~.\end{equation}
For a special choice $M={3\over 2}$ this model has enhanced $\CN=2$ supersymmetry as well as a $U(1)_R$ symmetry.

The various phases of the model are 

\begin{itemize}

\item $M>1/2$. Here for positive $\xi$ we have a gapped SUSY vacuum with $U(1)_2$ TQFT in the deep infrared.  For negative $\xi$ we have two trivial gapped SUSY vacua. The Witten index is constant as a function of $\xi$. 
\item $M<-1/2$. For positive $\xi$ we have a trivial gapped SUSY vacuum.  For negative $\xi$ we have a trivial gapped SUSY vacuum as well as a vacuum with $U(1)_2$ TQFT.  The Witten index is constant as a function of $\xi$
\item $|M|<1/2$. For positive $\xi$ we have two  gapped SUSY vacua, one trivial and one with TQFT $U(1)_2$.  For negative $\xi$ we have a trivial gapped SUSY vacuum. The Witten index is constant as a function of $\xi$

\end{itemize}

We will discuss the the lines $M=\pm\frac12$ momentarily. These are the walls where the Witten index jumps. Of course, these lines may take a different shape in the full theory, but the various phases and the topology of the phase diagram are robust. It is useful to again plot the various phases we have described above along with the Witten index of each of the gapped phases.

\begin{center}\includegraphics[scale=0.3]{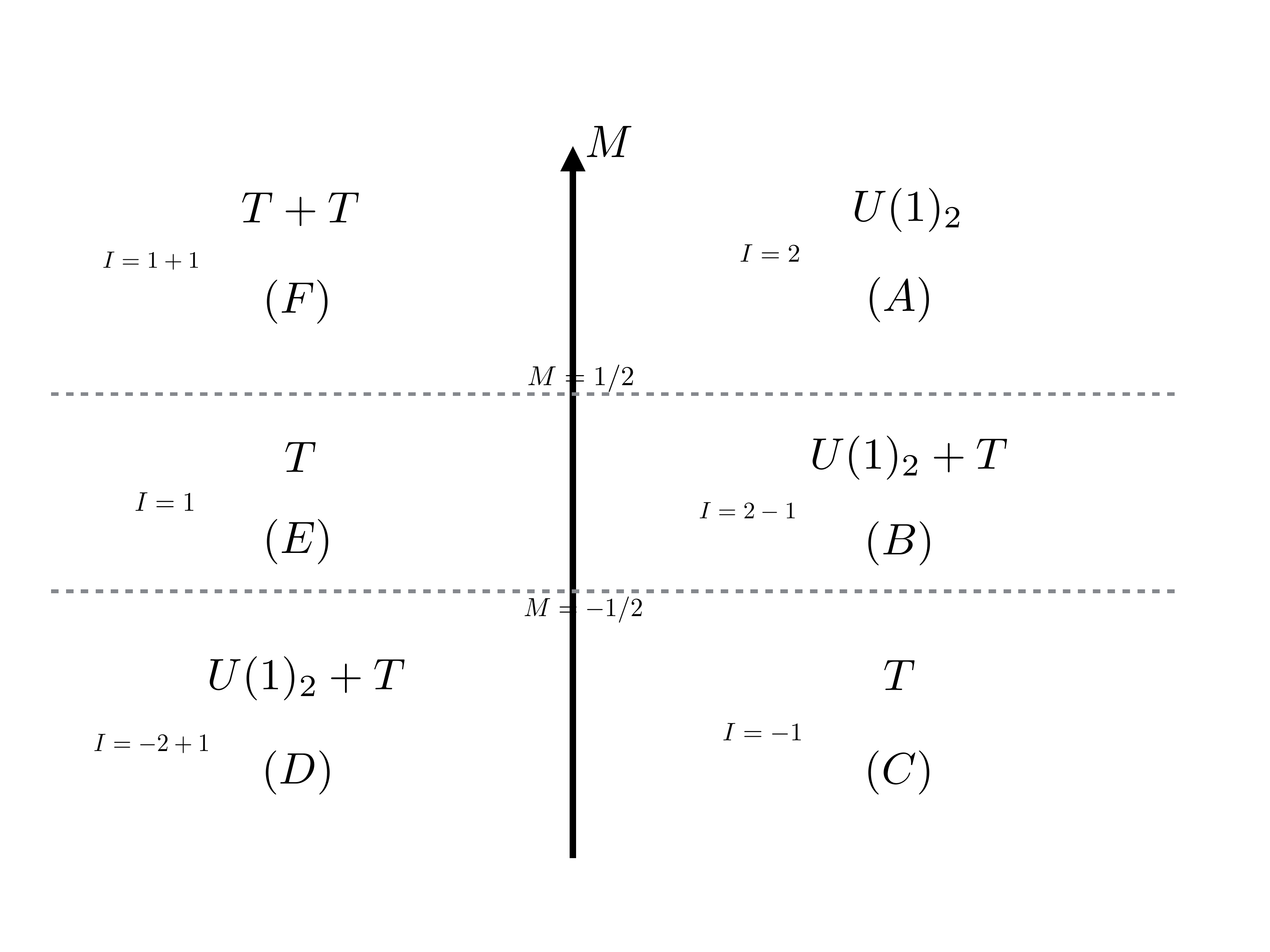}\end{center}

There are again 6 regions, labeled A,B,C,D,E,F clockwise and in each region we record the vacua and their Witten indices. The bold vertical line is at $\xi=0$, positive $\xi$ is to the right and negative $\xi$ is to the left. The dashed lines at $M=\pm\frac12$ are walls where the index must jump. The bold line at $\xi=0$ is the naive line of SCFTs. (These SCFTs do not have to be all distinct as we travel on the $M$ axis.)

Let us now make some observations -- except for $M=\pm1/2$, as we move horizontally on the diagram,
the Witten index does not jump. However, a SCFT must nonetheless occur somewhere around $\xi=0$ because the number of vacua and their low-energy properties change. In particular, at $\xi=0, M=3/4$ that SCFT has $\CN=2$ supersymmetry and it can studied in great detail.
As we move vertically on the diagram (away from $\xi=0$) the Witten index clearly jumps at the two walls at $M=\pm \frac12$. For example, when we move from region F to E a trivial SUSY vacuum disappears to infinity and as we proceed into region D a new SUSY vacuum with $U(1)_2$ TQFT  appears from infinity.

On the two walls at $M=\pm \frac12$ the superpotential is 
\begin{equation}\label{sup}W= gS|\tilde\Phi|^2\mp \frac14g^2S^2-\frac14g^2S|S|+\xi S=0~.\end{equation}
and the critical point equations are 
\begin{equation}\label{critical}S\tilde \Phi=0~,\quad g|\tilde \Phi|^2\mp \frac12g^2S-\frac12g^2|S|+\xi=0~.\end{equation}

On the $M=1/2$ wall the energy density for negative $S$ (and vanishing $\Phi$) is asymptotically constant and similarly on the $M=-1/2$ wall for positive $S$ (and vanishing $\Phi$) the energy density is constant. In both cases at $\xi=0$ there is a flat direction isomorphic to $\mathbb R_+$. This moduli space of vacua is not an artefact of the leading order approximation. The walls exist in the full theory and since there is a phase transition on the wall, the only way that the figure can be consistent is that a moduli space opens up at the SCFT on the wall. This is a general mechanism that can lead to moduli spaces of 
SUSY vacua even in the absence of a microscopic time reversal symmetry.

Note a remarkable fact: consider the point $M=-1/2, \xi=0$, at the intersection of the regions B,C,D,E. Around that point there is an emergent symmetry of reflections. The SCFT at $M=-1/2, \xi=0$ is therefore conjectured to have emergent time reversal symmetry. 
(As we will see, this is not the only point that has emergent time reversal symmetry in the infrared.)

Now simply tensor the theory of $S,\tilde \Phi$ coupled to a $U(1)_{3/2}$ gauge field by a $U(1)_2$ pure TQFT. 
Then, all the phases we found here can be seen to exactly coincide after the appropriate identification with the phases of the charge 2 particle coupled to a $U(1)_0$ gauge field.

In particular, the point $M=-1/2,\xi=0$ (tensored with a $U(1)_2$ TQFT) is dual to our $U(1)_0$ gauge field coupled to a charge 2 superfield with vanishing superpotential (i.e. $|\tilde M|=\infty$). Therefore, there is emergent time reversal symmetry at $M=-1/2,\xi=0$. 
In addition, we see that  $\tilde M=1$ (or equivalently $\tilde M=-1$) maps to $M=1/2$. The regions ABCDEF in the second figure therefore map to EFCDAB, respectively, in the first figure. 
Another interesting special case of this duality is the  map between $\tilde M=0$ and $M=3/4$. This leads to emergent time reversal symmetry at $M=3/4,\xi=0$. 
This latter case is a duality between two $\CN=2$ supersymmetric theories (of which one has emergent time reversal symmetry) and it can be subjected to stringent tests using the $\mathbb{S}^3$ and $\mathbb{S}^2\times\mathbb{S}^1$ partition functions.  We study this $\mathcal{N}=2$ duality in the appendix \ref{appendix:duality}.

It would be interesting to understand when the parameters $M,\tilde M$ are relevant. Clearly the SCFTs at $M=3/4\longleftrightarrow \tilde M=0$ and $M=-1/2\longleftrightarrow |\tilde M|=\infty$  are not the same since the former does not have a moduli space of 
vacua while the latter does. The former has ${\cal N}=1$ supersymmetry while the latter ${\cal N}=2$ supersymmetry.  Therefore, the duality we find here has both ${\cal N}=2$ and ${\cal N}=1$ versions. In both cases in one of the duality frames there 
is emergent time reversal symmetry. In the ${\cal N}=1$ version of the duality, the infrared SCFT has a moduli space of vacua isomorphic to $\mathbb R_+$.

Note that starting from the duality that we established above 
 $$U(1)_0+{\rm charge \ 2} \longleftrightarrow U(1)_2\otimes\left[U(1)_{3/2}+{\rm charge\ 1}\right]~,$$
 we can imagine deforming the theories by giving a mass to the bosons on the left-hand side and a corresponding mass to the fermion on the right-hand side. This leads to a non-supersymmetric duality between $U(1)_0+{\rm charge\ 2\ fermion}$ and $U(1)_2\otimes[U(1)_1+{\rm charge\ 1\ boson}]$. But since $U(1)_1+{\rm charge\ 1\ boson}$ is dual to a Dirac fermion, we recover precisely the duality between $U(1)_0+{\rm charge\ 2\ fermion}$ and $U(1)_2\otimes{\rm Dirac\ fermion}$, in agreement with~\cite{Cordova:2017kue,Gomis:2017ixy}. Both sides of the proposed duality have a ${\mathbb Z}_2$ 1-form symmetry with a 't Hooft anomaly. This anomaly is matched exactly as in the
 non-supersymmetric analogue of the duality. We refer to ~\cite{Cordova:2017kue,Gomis:2017ixy} for details and to~\cite{Kapustin:2014gua,Dierigl:2014xta,Gaiotto:2014kfa,Gaiotto:2017yup} (and references therein) for some more background on anomalies of one-form symmetries.

\section{Symmetry Enhancement in $\mathcal{N}=2$ QED and Supersymmetry Enhancement in a Wess-Zumino Model}\label{sec:sym_enhancement}
\medskip

Here we study a certain $\mathcal{N}=2$ duality that exhibits surprising features. On one side of the duality the global symmetry of the infrared theory is not manifest. On the other side of the duality, the supersymmetry of the infrared theory is not manifest.

We want to claim a duality between the following two theories:
\begin{enumerate}
\item A 3d ${\cal N}=2$ $U(1)$ gauge theory with vanishing Chern-Simons level and two chiral superfields of charge $1$. 
This theory has an unexpected IR enhancement of flavor symmetry to $SU(3)$. 
\item A 3d ${\cal N}=1$ Wess-Zumino model eight real chiral fields $\phi^a$ transforming in the adjoint of $SU(3)$,
and a real cubic superpotential 
\begin{equation}
W = \frac{1}{6}d_{abc} \phi^a \phi^b \phi^c~,
\end{equation}
\end{enumerate}

where $d_{abc} = 2\Tr[\{T_a,T_b \}T_c]$, with $T_a$, $a = 1,...,8$ the generators of $su(3)$, satisfying $\Tr[T^aT^b] = \frac12  \delta^{ab}$. This $\mathcal{N}=1$ theory is conjectured to have $\mathcal{N}=2$ supersymmetry (as well as a $U(1)_R$ symmetry) in the infrared. 

To support this surprising duality proposal, we will match the massive deformations of these theories, including the contact terms for background gauge fields. Additional arguments for this symmetry enhancement have also been suggested in \cite{Gang:2017lsr,Gang:2018wek}.
 
\subsection{Massive Deformations of the Wess-Zumino Model}

There is a simple mass deformation, with super-potential 
\begin{equation}\label{WZmodelNone}
W = \frac{1}{6}d_{abc} \phi^a \phi^b \phi^c + m_a \phi^a~.
\end{equation} 

The masses $m_a$ transform in the adjoint of the $SU(3)$ flavor symmetry. 
We can always conjugate them to the Cartan sub-algebra and to a Weyl chamber therein.
Generic masses will preserve a $U(1)^2$ symmetry, but there is a co-dimension $1$ 
locus where the preserved flavor symmetry enhances to $U(1) \times SU(2)$. 

To be concrete, we can collect the real scalars and masses into traceless Hermitian $3 \times 3$ matrices 
$\Phi = \phi_a T^a$ and $M = m_aT^a$. The superpotential becomes 
\begin{equation}
W= \frac23 \Tr \Phi^3 + \Tr M \Phi 
\end{equation}
and the equation for classical vacua is 
\begin{equation}
\Phi_*^2 + M = c 1_{3 \times 3}~.
\end{equation}
Observe that for $M=0$ (i.e. the undeformed model) there are no solutions other than the trivial one and hence 
there is no moduli space of vacua at the SCFT point. 

Taking  $M$ diagonal and generic, 
%
we see that  $\Phi_*$ is the square root of a diagonal matrix with positive entries and is also diagonal. Only two of the possible square 
roots are traceless, giving us two semi-classical vacua with unbroken $U(1)^2$ global symmetry. The two vacua spontaneously 
break time-reversal symmetry, as $\Phi$ is a pseudo-scalar. 

On the other hand, if $M$ is special and preserves $U(1) \times SU(2)$,
say, diagonal with entries $(m,m,-2m)$, then we have two possibilities:
\begin{enumerate}
\item For $m<0$, $c=-2 m$ and $\Phi_*$ has a traceless $2 \times 2$ block which is a square root of $1_{2 \times 2}$.
That gives an $\C P^1$ worth of vacua, spontaneously breaking $SU(2) \times U(1) \to U(1)^2$.
\item For $m>0$, $c=2 m$ and there are again two vacua, which preserve $SU(2) \times U(1)$. These two vacua are related by time reversal symmetry.
\end{enumerate}
In the $(M_{11}, M_{22})$ plane, we thus have three half-lines with an $\C P^1$ worth of vacua, and $2$ vacua that are related by time reversal symmetry elsewhere. Time reversal symmetry under which $\Phi$ is a pseudo-scalar acts by the antipodal map on the $\C P^1$. The analysis of these massive deformations is also done in detail (in 
components) in appendix A. The phase diagram is shown in Fig. \ref{fig:threelines}.

As $\C P^1$ is a K\"ahler manifold, the low energy non-linear sigma model on these three lines where the global $SU(3)$ symmetry is explicitly broken to $SU(2)\times U(1)$ 
gains an ${\cal N}=2$ supersymmetry in the deep infrared. 
(In other words, the  $\C P^1$ model with a round metric and at most two derivative interactions with ${\cal N}=1$ supersymmetry must in fact have ${\cal N}=2$ supersymmetry as well as a $U(1)_R$ symmetry.) 

We would like to argue that this supersymmetry enhancement is also a property of the IR SCFT at the origin of parameter space (i.e. at $M=0$). The new infrared supercharge starts its life as a cubic  in the fermions (the spins are symmetrized so the flavor indices must be anti-symmetrized and hence we have to use the $f_{abc}$ symbol of $su(3)$) and thus the prediction is that the dimension renormalizes down from $3$ in the ultraviolet to $2.5$ in the infrared. 
\begin{figure}[h]
\centering
\includegraphics[width=0.6\linewidth]{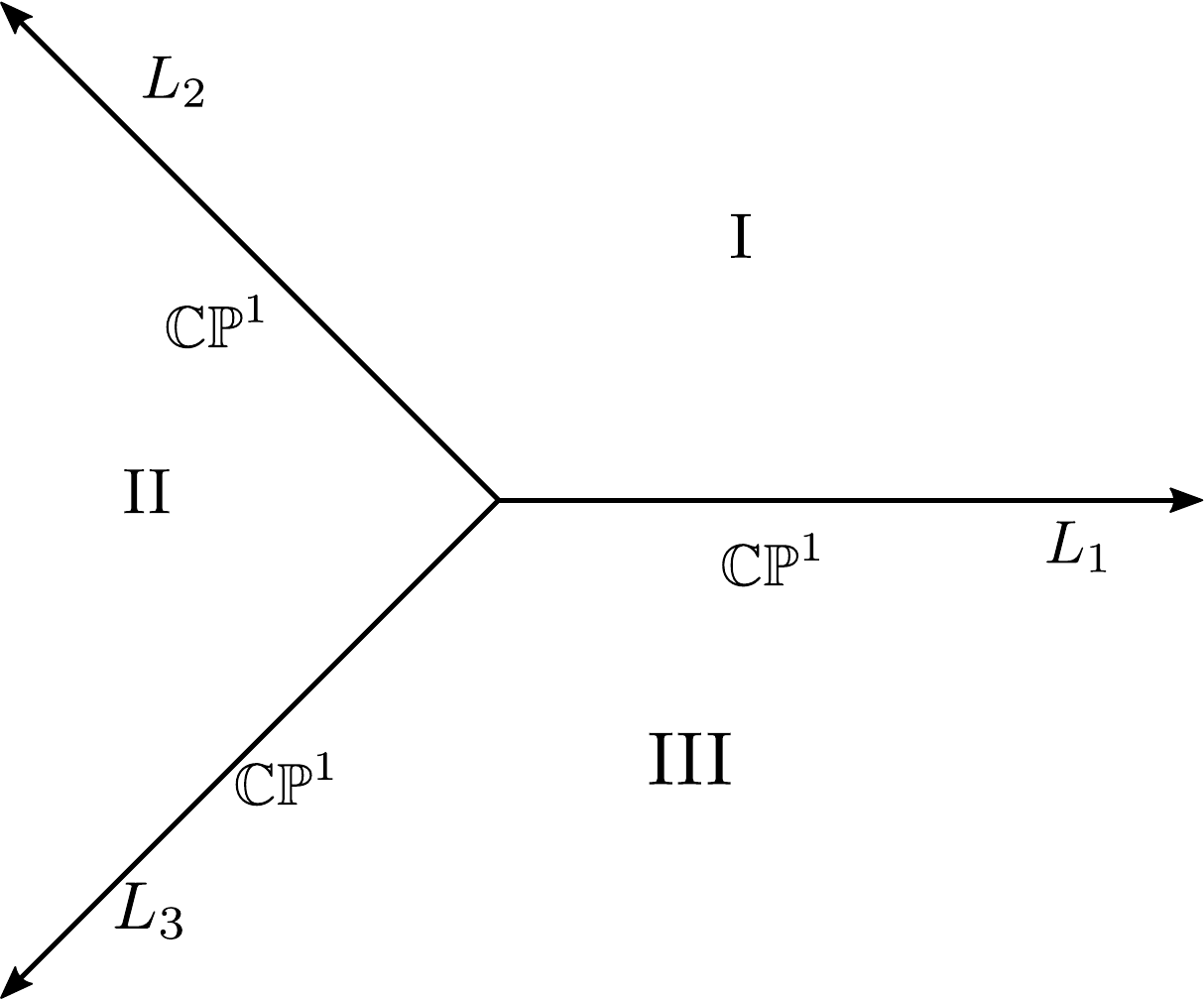}
\caption{Phase diagram of the Wess-Zumino model and gauge theory. For the Wess-Zumino model, this is ($M_{11}$,$M_{22}$) plane and three half lines are given by $L_1: M_{11} -M_{22} = 0\ (M_{11}<0)$, $L_2: 2M_{11}+ M_{22}=0\ (M_{11}<0)$, and $L_3 : M_{11}+2M_{22} = 0 \ (M_{11}>0)$. For the gauge theory this is ($t$,$m_f$) plane and three lines are given by $L_1: m_f = 0\ (t>0)$, $L_2: t+m_f = 0\ (m_f>0) $, $L_3: t-m_f = 0\ (m_f<0) $. On the three solid half lines, we have an $SU(2)\times U(1)$ preserving deformation, and $\mathbb{C}\mathbb{P}^1$ worth of vacua due to spontaneous symmetry breaking. Otherwise, we have 2 isolated vacua due to time reversal symmetry breaking. }
\label{fig:threelines}
\end{figure}

It will prove useful to analyze the background Chern-Simons couplings in the massive vacua. 
The superpotential expanded to quadratic order around the vev $\Phi_*$ takes the form 
\begin{equation}
\delta W = \Tr \Phi_* \delta \Phi^2~.
\end{equation}
As the vacuum vev $\Phi_*$ is diagonal, the coefficient for the off-diagonal $|\delta \Phi_i^j|^2$ is simply $(\Phi_*)_i^i + (\Phi_*)_j^j = -(\Phi_*)_k^k$,
where $k$ is the index different from $i$ and $j$. 
Since $\Phi_*$ is traceless, we either have two positive masses and one negative, or two negative masses and one positive. As the two vacua have opposite 
$\Phi_*$, the pattern of masses is opposite in the two vacua and so are the background Chern-Simons couplings. 
This is compatible with the spontaneous breaking of time-reversal symmetry. 

If we denote the $U(1)^2$ background gauge connection as $A_i$, with $A_1 + A_2 + A_3=0$, 
then depending on which of the three chambers we are in the vacua will have background CS couplings 
$\pm A_1 d A_2$, $\pm A_2 d A_3$ or $\pm A_1 d A_3$. 

If we sit on a line with unbroken $U(1) \times SU(2)$, we choose a parametrization of $\Phi$ as
\begin{equation}
\Phi=\left(
\begin{array}{ccc}
\phi_8 +R_3 & R_1+i R_2 & X \\
R_1-i R_2 & \phi_8-R_3 & \bar{Y} \\
\bar{X} & Y & -2 \phi_8 \\
\end{array}
\right);
\end{equation}
where $R_i$ and $\phi_8$ are real and $X$, $Y$, and $Z$ are complex. Plugging the expectation value $\Phi_* = \text{diag}\{m,m,-2m \}$ with $m>0$, the mass terms can be written as
\begin{equation}
2m(R_1^2 + R_2^2 + R_3^2) -m(|X|^2+|Y|^2) - 6 \phi_8^2
\end{equation}
In other words, we have an $SU(2)$ doublet of fields with $U(1)$ charge $1$ and mass $-m$, a triplet of $SU(2)$ with $U(1)$ charge  $0$ and mass $2m$, and a singlet $\phi_8$.

A background Chern-Simons term for $SU(2)$ induced by integrating out matter field with mass $m_i$ in the $R_i$ representation. This is given by \cite{Intriligator:2013lca}
\begin{equation}
k_{SU(2),eff} = k_{SU(2),bare} + \frac{1}{2} \sum_{i} T(R_i) \ \text{sign}(m_i)
\end{equation}
where $T(R_i)$ is the quadratic index of representation $R_i$ defined by
\begin{equation}
\Tr[T^a_RT^b_R] = \frac12 T(R) \delta^{ab}
\end{equation}
with the normalization $T(\mathbf{2}) = 1$, $T(\mathbf{3}) = 2$. Therefore, \begin{equation}
k_{SU(2),eff}  = \frac12~. \label{eq:kSU(2)LG}
\end{equation}
(We will soon compare this do the dual gauge theory.)
\subsection{Massive Deformations of the ${\cal N}=2$ Gauge Theory}
We denote the two chiral superfields with the gauge charge $+1$ as $Q$ and $\tilde{Q}$ and the bottom component in the vector multiplet $V$ is denoted as $s$. The theory has four natural (i.e. visible in the microscopic theory) ${\cal N}=2$-preserving real mass deformations, associated to the $U(1)_t \times SU(2)_f$ flavor symmetry.
Without loss of generality, we can consider the two mass deformations associated to the Cartan subalgebra. We will denote the generator associated to the Cartan of $SU(2)$ by  $m_f$ and the FI parameter for the topological 
$U(1)_T$ symmetry is denoted by $t$. The supersymmetric vacua can be found by looking for the critical points of the one-loop corrected potential, as in~\cite{Intriligator:2013lca}. The equations are thus
\begin{align}
|Q|^2 + |\tilde{Q}|^2 &= t_{eff}\\
(s + m_f)Q &= 0\\
(s - m_f)\tilde{Q} & = 0
\end{align}
where the one-loop effective FI parameter is 
\begin{equation}
t_{eff} = t + \frac12 (|s + m_f| + |s-m_f|)~.
\end{equation}

For $t=m_f=0$ the theory has no moduli space of vacua, as our Wess-Zumino model. 
When $t>0$, $m_f=0$, the theory has a $\C P^1$ of vacua parameterized by vevs of the chiral fields (and $s=0$). 
This corresponds to the half line $L_1$ in Fig.\ref{fig:threelines}.
Turning on $m_f$ reduces the vacua to the two poles of $\C P^1$. These two vacua are related by time reversal symmetry since $s=\pm m_f$ in these two vacua. Hence, time reversal symmetry is spontaneously broken in this gapped phase.

When $t<0$, something special happens when $m_f = \pm t$, i.e. on the half lines $L_2$ and $L_3$: a Coulomb branch parameterized by the 
real scalar in the vectormultiplet, with $t \leq s \leq -t$ exists. In this phase $Q=\tilde Q=0$.
The two chirals have opposite 
masses and contribute $\frac12(s-t)- \frac12(s+t) = -t$ to the effective FI parameter.
The real scalar combines with the dual photon to give a new $\C P^1$. 

For intermediate values of $m_f$ in the region $\mathrm{II}$ in Fig.\ref{fig:threelines}, we have vacua where both chirals have either positive or negative mass and no vacuum expectation values. Therefore the effective FI parameter is $t \pm s$, so that $s$ is fixed either to $s =t$ or to $s = -t$. We have again two vacua and time reversal symmetry is spontaneously broken.

The $S_3$ symmetry of the phase diagram, permuting the three $\C P^1$ half-lines, 
is clearly suggestive of an enhancement of the flavour symmetry to $SU(3)$. 

This conjectural $S_3$ Weyl symmetry can be realized as a group of self-dualities of the theory. 
The simplest way is to use the ${\cal N}=2$ particle-vortex duality to convert a chiral multiplet to 
a $U(1)_{\frac12}$ theory coupled to a dual chiral multiplet. The resulting theory has 
a new manifest $\mathbb{Z}_2$ symmetry exchanging at the same time the two gauge fields and the two chiral multiplets.
Together with the Weyl symmetry of $SU(2)_f$, this generates the required $S_3$. \footnote{The Weyl symmetry can be made manifest in both the index and $S_b^3$ partition functions of the theory, by manipulations analogue to that particle-vortex duality. See appendix \ref{appendix:duality}}

The massive phase that we have found with $m_f=0$ for $t<0$ preserves the global $SU(2) \times U(1)$ symmetry. In this case the doublet $(Q,\tilde{Q})$ (in one of the two vacua which are related by time reversal symmetry) acquires a positive mass, generating a background $SU(2)$ Chern-Simons term after it is integrated out 
\begin{equation}
k_{SU(2),eff} = \frac12 
\end{equation}
This matches the result \eqref{eq:kSU(2)LG} in the Wess-Zumino model.

We can look at the background Chern-Simons coupling in the massive vacua. With generic $m_f$ and $t$, $U(1)_t$ and $U(1)_f \subset SU(2)$ are preserved. Again, integrating out fermions with real mass $m_i$ will induce background Chern Simons terms given by
\begin{equation}
k_{ab,\text{eff}} = k_{ab,\text{bare}} + \frac12 \sum_{i} n_{a,i} n_{b,i} \ \text{sign}(M_i)
\end{equation}
where $n_{a,i}$ is the charge of $i$-th fermion under $U(1)_a$, $M_i = m_i + n_{g, i} s$ is the effective mass, with gauge charge $n_{g,i}$. In the region $\mathrm{I}$, the result for two vacua is 

\begin{equation}
\left(\begin{array}{cc}
k_{JJ} & k_{Jf} \\ 
k_{fJ} & k_{ff}
\end{array} \right)_{\mathrm{I}}= \qquad
\left(\begin{array}{cc}
	0 & 1 \\ 
	1 & 2
\end{array} \right), \qquad
\left(\begin{array}{cc}
0 & -1 \\ 
-1 & -2
\end{array} \right) \label{eq:kU(1)gauge1}
\end{equation}
In the region $\mathrm{III}$, the situation is similar 
\begin{equation}
\left(\begin{array}{cc}
k_{JJ} & k_{Jf} \\ 
k_{fJ} & k_{ff}
\end{array} \right)_{\mathrm{III}}= \qquad
\left(\begin{array}{cc}
0 & -1 \\ 
-1 & 2
\end{array} \right), \qquad
\left(\begin{array}{cc}
0 & 1 \\ 
1 & -2
\end{array} \right) \label{eq:kU(1)gauge3}
\end{equation}

Region $\mathrm{II}$ is a bit different since the signs of masses of $Q$ and $\tilde{Q}$ are the same and $s$ is equal to FI parameter $\pm t$, which forces gauge field to identify with $A_J$.\footnote{There is also a more physical way to look at it. In the positive mass vacuum, integrating out $Q$ and $\tilde{Q}$ generates level one Chern Simons term $\frac{1}{4\pi} ada$ for the gauge field $a$. Then $\frac{1}{4\pi} (ada + 2adA_J)$ can be written as $\frac{1}{4\pi}( (a-A_J)d(a-A_J) - A_JdA_J)$. The first term is an trivial theory $U(1)_1$\cite{Seiberg:2016gmd} and we have $k_{JJ} = -1$. Likewise we have $k_{JJ} =1$ for the other vacuum.
	
} We then have nonzero $k_{JJ}$.
\begin{equation}
\left(\begin{array}{cc}
k_{JJ} & k_{Jf} \\ 
k_{fJ} & k_{ff}
\end{array} \right)_{\mathrm{II}}= \qquad
\left(\begin{array}{cc}
1 & 0 \\ 
0 & -1
\end{array} \right), \qquad
\left(\begin{array}{cc}
-1 & 0 \\ 
0 & 1
\end{array} \right) \label{eq:kU(1)gauge2}
\end{equation}
 
On the other hand, we can also compute the ${\bf k}$ matrix in the three regions on $(M_{11},M_{22})$ plane of Wess-Zumino model side. The charges of the different fields under the global symmetry are given by
\begin{equation}
\begin{tabular}{cccccccc}
	\hline 
	& Q & $\tilde{Q}$ & X & Y & Z & $\phi_3 $ & $\phi_8$ \\ 
	\hline 
	$U(1)_f$ & 1 & -1 & 1 & 1 & -2 & 0 & 0 \\ 
	\hline 
	$U(1)_J$ & 0 & 0 & 1 & -1 & 0 & 0 & 0 \\ 
	\hline 
\end{tabular} 
\end{equation}
\begin{center}
	\begin{tabular}{cccc}
	\hline 
	& $X$ & $Y$ & $Z$ \\ 
	\hline 
 $A_1$ & $1$ & $0$ & $1$ \\ 
	\hline 
 $A_2$ & $1$ & $1$ & $0$ \\ 
	\hline 
$A_3$ & $0$ & $1$ & $1$ \\ 
	\hline 
\end{tabular}
\end{center} 
where we rewrite $\Phi = \phi_a T^a$ in the Chevalley basis. In particular, $X = \frac12 (\phi_4 -i\phi_5)$, $ Y = \frac12 (\phi_6+i\phi_7)$, $Z = \frac12(\phi_1 + i\phi_2)$. The masses of the three complex fields $X$, $Y$ and $Z$ are given by the Hessian matrix of the scalar potential at the vacua. A simple calculation gives us the same ${\bf k}$ matrix \eqref{eq:kU(1)gauge1},\eqref{eq:kU(1)gauge2},\eqref{eq:kU(1)gauge3}, as the gauge theory in the three regions respectively. Upon identifying $A_f = \frac12 A_2$ and $A_J = A_1+\frac12 A_2$ in the previous section, we get $\pm A_2dA_3$, $\pm A_3dA_1$, and $\pm A_1dA_2$ in the region $\mathrm{I}$, $\mathrm{II}$, $\mathrm{III}$ respectively. This serves as another consistency check of our duality. 

\section{An ${\cal N}=1$ Duality Between SQED and a Wess-Zumino Model}
Here we would like to make some comments about an ${\cal N}=1$ duality which can be derived from our 
${\cal N}=2$ duality above. 
We will see that this ${\cal N}=1$ duality  is also closely related to the  
$\text{SQED}_1$-XYZ duality \cite{Aharony:1997bx}.

The dual pair is  
\begin{enumerate}
\item A 3d ${\cal N}=1$ $U(1)$ gauge theory with two chirals of charge $1$, total Chern-Simons level $0$.
This theory has an $SU(2)_f \times U(1)_t$ global symmetry. 
\item A 3d ${\cal N}=1$ Wess-Zumino model with seven real chiral fields: a complex  $SU(2)_f$ doublet $u^\alpha$ 
of $U(1)_t$ charge $1$ and a real  $SU(2)_f$ triplet $R_{\alpha \beta}$. We also add a 
real cubic superpotential 
\begin{equation}\label{cubicsup}
W = R_{\alpha \beta} u^\alpha \bar u^\beta~.
\end{equation}
We denote $\bar u^\beta = (u^{\beta})^\ast$ (i.e. it is just the complex conjugate superfield) and indices are raised or lowered using the $SU(2)$ invariant tensor $\epsilon^{\alpha \beta}$. 
\end{enumerate}

We will first motivate the duality and then connect it to the duality we have presented in the previous section. This ${\cal N}=1$ duality has appeared before in a different context, for instance,~\cite{Gremm:1999su,Gukov:2002es}.

First let us compare the moduli spaces of vacua of the two theories before we add any deformations. 
We begin with the $U(1)_0$ with two superfields  $\Phi_1,\Phi_2$ carrying charge 1 under the $U(1)$ gauge symmetry. 
We have a classical moduli space of vacua, where the gauge symmetry is broken (everywhere except at the origin). We can parameterize the moduli space by the expectation values of $\Phi_1$ and $\Phi_2$ while removing one overall phase which is gauged.
Therefore we can parameterize the moduli space by $|\Phi_1|,|\Phi_2|$, and $\arg\left(\frac{\Phi_2}{\Phi_1}\right)$. The model has a global $SU(2)$ symmetry which has a $U(1)$ subgroup that acts by shifting $\arg\left(\frac{\Phi_2}{\Phi_1}\right)$ while leaving $|\Phi_1|,|\Phi_2|$ intact. Therefore, considering now the full theory and not just the classical theory, we see that the superpotential cannot depend on  $\arg\left(\frac{\Phi_2}{\Phi_1}\right)$. And since 
$|\Phi_1|,|\Phi_2|$ are time reversal even, it cannot depend on them either. Therefore the classical moduli space of vacua is not lifted. 
\begin{equation}\label{modulispacetwo}{\mathcal M}_{vac}=\mathbb{R}^3~.\end{equation}
At a generic point on the moduli space the $SU(2)$ global symmetry is broken to $U(1)$ and hence we can equivalently parameterize the moduli space by the expectation value of $|\Phi_1|^2+|\Phi_2|^2$ and the $\mathbb{S}^2={SU(2)\over U(1)}$ worth of Nambu-Goldstone vacua fibered over it. 

Now let us consider the moduli space of vacua of the Wess-Zumino model~\eqref{cubicsup}. There is classically a 
moduli space parameterized by $R_{\alpha\beta}$. Going far on this moduli space, we can integrate out $F,\bar F$ and write an effective superpotential $W_{eff}=W_{eff}(R_{\alpha\beta})$. This effective superpotential is constrained by $SU(2)$ invariance and by time reversal symmetry, which acts by $R\to -R$. This leaves arbitrary 
$SU(2)$ invariant terms with an odd number of $R$ fields. It is easy to see that all such terms vanish identically
by simply diagonalizing $R$.  Therefore, the moduli space is spanned by $R_{ab}$ and hence we get ${\mathcal M}_{vac}=\mathbb{R}^3$, as in the Wess-Zumino model. 

Now we compare some deformations of the model, starting from the non-$SU(2)$ invariant mass deformation. 
Without loss of generality we consider 
$$W=m|\Phi_1|^2-m|\Phi_2|^2~.$$
This mass deformation breaks the global $SU(2)$ symmetry explicitly to $U(1)$.
At nonzero positive $m$ we have an effective theory of a $U(1)$ gauge field without a Chern-Simons term. The matter fields are massive. Therefore, it can be dualized to a real compact superfield and hence we have a circle of supersymmetric vacua. Similarly, for negative $m$ we have a circle of supersymmetric vacua. In both cases, these circles of supersymmetric vacua can be interpreted as due to the spontaneous breaking of $U(1)_t$.

The dual description of this mass deformation corresponds to adding a linear term in $R$, which we can take without loss of generality to be $\delta W =m R_{11}$ (remember that $R_{11}=-R_{22}$). The equations of motion now take
the form 
\begin{eqnarray*}
&|u^1|^2-|u^2|^2=m~,\\
&u^1\bar u^2=0~,\\
&R_{\alpha\beta}u^\alpha=0~.
\end{eqnarray*} 
The solution is clearly $R=0$, and for positive $m$ we have nonzero $u^2=0$, $|u^1|^2=m$, and for negative $m$ we have $R=0$, $u^1=0$, $|u^2|^2=-m$. So in both cases we have a circle of vacua with a spontaneously broken $U(1)_t$ symmetry. 

Finally, we can have an $SU(2)$ invariant mass deformation. In the Abelian gauge theory this corresponds to 
$$W=m|\Phi_1|^2+m|\Phi_2|^2~.$$
At positive $m$ we have a trivial supersymmetric vacuum (here we use the fact that $U(1)_1$ is a trivial TQFT) and for negative $m$ we have likewise another trivial supersymmetric vacuum. (The Witten index at positive $m$ is +1 and at negative $m$ it is $-1$.) 
In the Wess Zumino model, this deformation is mapped to a deformation of the superpotential by 
\begin{equation}\label{invsupdef}\delta W=m \eta_{\alpha \beta} u^\alpha\bar u^\beta+c_2 R_{\alpha\beta}R^{\alpha\beta}~,\end{equation}
where $\eta_{\alpha\beta}$ is as usual the identity matrix.\footnote{Note that the other $SU(2)$ singlet deformation, $u^\alpha\bar u_\alpha$, is redundant (in the sense that it can be removed by a change of variables).} The coefficient 
$c_2$ is unknown.  
 Classically, with the deformation~\eqref{invsupdef} the $R$ equations of motion still set $u=0$ and due to $c_2$ 
 $R$ is pinned to the origin. (Even if $c_2=0$, while $R$ is classically arbitrary, there will be an effective potential on the moduli space parameterized by $R$ since there is no more time reversal symmetry.) As long as $R$ is pinned 
 to the origin (either due to $c_2$ or due to a radiatively generated potential) we get a massive trivial supersymmetric vacuum with unbroken $U(1)_t\times SU(2)$, as required by the duality.

Therefore, under the duality, $R_{\alpha \beta}$ maps to a triplet of mesons and the singlet meson maps roughly to $|u|^2$ (with a possible admixture of $R^2$).

This pair of ${\cal N}=1$ theories is related by simple ``flip'' operations both to the dual pair in the previous section 
and to the SQED$_1$-XYZ mirror symmetry. It is also related to the basic 3d ${\cal N}=4$ mirror pair~\cite{Aharony:1997bx,Kapustin:1999ha}. 

In order to see the relation to the former, we can ``flip'' the real $U(1)_t$ moment map operator 
in the 3d ${\cal N}=2$ $U(1)$ gauge theory.  Flipping means adding a new real multiplet $R$
with linear superpotential coupling to $S$ ($S$ is the real ${\cal N}=1$ superfield containing in the bottom component the real scalar of the vector multiplet), i.e. promoting the ${\cal N}=2$ FI parameter to a dynamical real 
superfield. This clearly preserves ${\cal N}=1$ supersymmetry.
This new term allows to integrate out $R$ and $S$. No extra ${\cal N}=1$ superpotential terms can be induced, 
as no gauge-invariant operators odd under time reversal symmetry (and invariant under the global symmetries ) are available. We thus have the 
${\cal N}=1$ $U(1)$ gauge theory with two chirals of charge $1$ at low energies.

On the Wess-Zumino side of the duality, the flip breaks $SU(3)$ to $SU(2)_f \times U(1)_t$. The 8 real chiral multiplets 
decompose into a complex  $SU(2)_f$ doublet $u^\alpha$ 
of $U(1)_t$ charge $1$, a real $SU(2)_f$ triplet $R_{\alpha \beta}$ and a singlet, to be identified with $S$. 
The flipping operation again removes $S$ and leaves a Wess-Zumino model of seven real chiral operators and cubic superpotential 
\begin{equation}\label{flipresult}
W = R_{\alpha \beta} u^\alpha \bar u^\beta~,
\end{equation}
which is the model we have studied here. 
Again, no extra terms compatible with the symmetries are available other than~\eqref{flipresult}.

Now we will explain the relation of this ${\cal N}=1$ duality with the familiar ${\cal N}=2$ mirror symmetry. We again start from the gauge theory side (${\cal N}=2$ SQED$_1$) and flip the real $U(1)_t$ moment map operator. 
The result is, again, the ${\cal N}=1$ $U(1)$ gauge theory with two multiplets of charge $1$.
No superpotential can be generated. Notice the enhancement of flavor symmetry from 
$U(1)^2$ to $SU(2)_f \times U(1)_t$. This enhancement is due to the fact that in ${\cal N}=1$ theories, there is no 
difference between charge 1 and charge -1 multiplets.

On the $XYZ$ side, the operator we are flipping is the real moment map 
$|X|^2 - |Y|^2$. The real superpotential is deformed from $ \mathrm{Re} (XYZ)$ to 
\begin{equation}
W = \mathrm{Re} Z (X Y + \bar X \bar Y) + i \mathrm{Im} Z (X Y - \bar X \bar Y) + R (|X|^2 - |Y|^2)
\end{equation}
If we form a doublet $u^\alpha =  (\bar X, Y)$, then the superpotential becomes  
\begin{equation}
W = \mathrm{Re} Z (u^2 \bar u^1 + u^1 \bar u^2) + i \mathrm{Im} Z (u^2 \bar u^1 - u^1 \bar u^2) + R (u^1 \bar u^1 - u^2 \bar u^2) = R_{\alpha\beta}u^\alpha\bar{u}^\beta
\end{equation}
which has enhanced $SU(2)_f \times U(1)_t$ global symmetry, with a triplet $R_{\alpha \beta} = (\mathrm{Re} Z, \mathrm{Im} Z , R)$. This is precisely our dual ${\cal N}=1$ Wess-Zumino model.

What we have shown is that the {\cal N}=1 duality described here can be derived either from the standard mirror symmetry or starting from our new duality in the previous section. We could have of course derived this duality direction from the 3d ${\cal N}=4$ mirror symmetry between SQED$_1$ and a Free twisted hyper. 
We can do so by an ${\cal N}=1$ $S$ operation, coupling a $U(1)$ ${\cal N}=1$ gauge field to the free twisted hypermultiplet, 
in such a way that it ``ungauges'' the $U(1)$ gauge field on the SQED$_1$ side. 

The Wess-Zumino model real superpotential is indeed the ${\cal N}=1$ description of the coupling between the three real scalars $R_{\alpha \beta}$ 
in the ${\cal N}=4$ gauge multiplet of SQED$_1$ and the hypermultiplet flavors $u^\alpha$.

Alternatively, we can ``flip'' all three real moment map operators for $U(1)_t$ in the ${\cal N}=4$ mirror pair. The result is the same. To summarize, the dualities we have considered are all connected by the 'flip' operation as in the figure below.
\begin{figure}[h]
\centering
\includegraphics[width=0.8\linewidth]{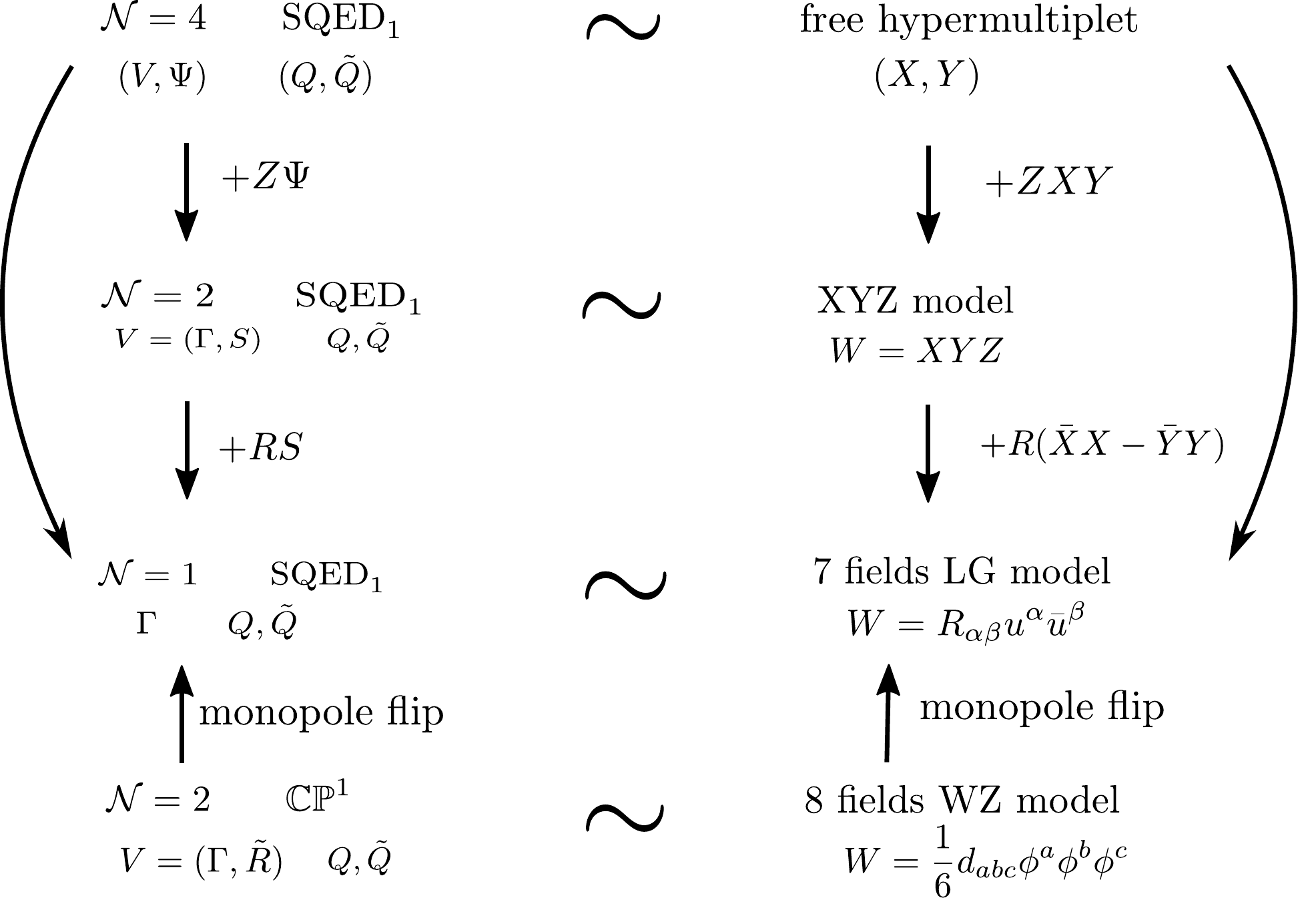}
\label{fig:web}
\end{figure}

\medskip

\section{${\cal N}=1$ Supersymmetric $SU(N)_0$ Gauge Theory with $N_f$ Quarks}

An interesting class of time-reversal invariant non-Abelian gauge theories is given by $SU(N)_0$ gauge theory (the subscript indicates the quantum Chern-Simons level) coupled to $N_f$ multiplets of quarks $Q_i$ in the fundamental representation ($i=1,...,N_f$). The number of fundamental representations is constrained such that 
\begin{equation}\label{conscond}N=N_f \ {\rm mod} \ 2~.\end{equation} 
This is a necessary and sufficient condition for the time reversal invariant theory to exist. (We will see below that if this condition is not obeyed one finds various nonsensical results.)

Suppose the superpotential vanishes classically. Then the theory has time reversal symmetry along with $U(N_f)$ global symmetry.\footnote{We ignore various discrete identifications.}
In fact, if we require time reversal symmetry and $U(N_f)$ symmetry, it is not possible to write any superpotential which is a function of the $Q_i$. Therefore, to all orders in perturbation theory, the renormalization group (around the ultraviolet, where the $Q_i$ are good degrees of freedom) does not generate a superpotential if we take the superpotential to vanish at tree level.

That means that the large moduli space of supersymmetric ground states that exists in the classically massless theory persists to all orders in perturbation theory.

Let us assume that $N_f<N_c$. In that case we can parameterize the moduli space by the expectation values of the mesons 
\begin{equation}\label{mesons} M_{ij}=Q_i^\dagger Q_j~.\end{equation}
To see that, we can use the gauge symmetry and $U(N_f)$ symmetry to bring the $N\times N_f$ matrix of $Q_i$ to the form 
\begin{equation}\label{genVEV}Q=\left(\begin{matrix} a_1 & 0 & ... & 0 \\ 0 & a_2 & ... & 0 \\ ... & ... & ... & ...\\ 0 & 0 & ... & a_{N_f}\\ 0 & 0 & ... & 0\cr ... & ... & ... & ...\\ 0 & 0 & ... & 0 \end{matrix}\right)\end{equation}
with $a_i\geq 0$. For generic $a_i$ the global symmetry is broken as 
\begin{equation} U(N_f)\to U(1)^{N_f}~.\end{equation}
(And the gauge symmetry is broken as $SU(N)\to SU(N-N_f)$.) Acting on~\eqref{genVEV} with the broken $U(N_f)\over U(1)^{N_f}$ generators we therefore obtain a $N_f^2-N_f+N_f=N_f^2$ real-dimensional moduli space. This space is parameterized by the $N_f^2$ mesons~\eqref{mesons}. (The mesons can be thought of as a Hermitian $N_f\times N_f$ matrix.)

The degrees of freedom on the moduli space to all orders in perturbation theory are therefore these $N_f^2$ massless mesons but, importantly, (at a generic point on moduli space) also the $SU(N-N_f)_0$ vector multiplet. 
The two sectors are not entirely decoupled. The vector multiplet effective action to first approximation is the canonical one, independent of the $M_{ij}$ but there are some irrelevant operators tying the two sectors together. It is easy to write the leading terms that couple the two sectors, but we would not need them here.

Thus far the analysis was to all orders in perturbation theory. However, non-perturbatively, the pure $SU(N-N_f)_0$ vector multiplet theory breaks supersymmetry~\cite{Witten:1999ds}. In the infrared one has a Goldstino along with the (spin) time reversal invariant
$U({N-N_f\over 2})_{{N-N_f\over 2},N-N_f}$ TQFT~\cite{Gomis:2017ixy} (this TQFT makes sense due to~\eqref{conscond}). The vacuum energy is set by the gauge coupling. 
Therefore, non-perturbatively, our $N_f^2$ real-dimensional moduli space is lifted. 

Since the coupling between the mesons and the vector multiplet vanishes for $|M|\to \infty$, we see that the vacuum energy is asymptotically constant. Unlike in the analogous theories in four dimensions (with four supercharges)~\cite{Affleck:1983mk}, there is no supersymmetric ground state at infinity. 

Note that the case $N_f=N-1$ does not exist in the sense that it cannot preserve time reversal symmetry~\eqref{conscond}. Hence our analysis is in fact valid for all $N_f<N_c$ as there is always a nontrivial unbroken gauge group (which in turn leads to dynamical supersymmetry breaking and lifts the moduli space non-perturbatively).

The cases $N_f\geq N_c$ are more interesting as a generic point on the moduli space breaks the gauge symmetry completely. Therefore, there is no mechanism to lift the moduli space at a generic point and hence there would be a real moduli space of SUSY vacua in the full theory. We leave the analysis of $N_f\geq N_c$ for the future.

\medskip

\section*{Acknowledgements}

We thank O.~Aharony, J.~Gomis, S.~Razamat,  M.~Rocek, N.~Seiberg, A.~Sharon, and R.~Yacoby. The research of D.G and J.W. was supported by the Perimeter Institute for Theoretical Physics. Research at Perimeter Institute is supported by the Government of Canada through Industry Canada and by the Province of Ontario through the Ministry of Economic Development and Innovation. Z.K. is supported in part by an Israel Science Foundation center
for excellence grant and by the Simons Foundation grant 488657 (Simons Collaboration
on the Non-Perturbative Bootstrap)

\appendix
\section{Detailed Analysis of the Wess-Zumino Model~\eqref{WZmodelNone}}
We first write the superpotential explicitly, using the $d$-symbols of $SU(3)$, in the  basis of Gell-Mann matrices:

\begin{eqnarray*}
&W&={1\over 6}\biggl[ \sqrt 3\phi^8 \left((\phi^1)^2+(\phi^2)^2+(\phi^3)^2\right) \\ 
&+&{3\over 2}\left(  2 \phi^1\phi^4\phi^6+2\phi^1\phi^5\phi^7+2\phi^2\phi^5\phi^6+\phi^3(\phi^4)^2+\phi^3(\phi^5)^2 \right) \\ 
&-&{3\over 2} \left(2\phi^2\phi^4\phi^7+\phi^3(\phi^6)^2+\phi^3(\phi^7)^2\right) \\ 
&-&{\sqrt 3\over 2} \phi^8 \left((\phi^4)^2+(\phi^5)^2+(\phi^6)^2+(\phi^7)^2\right)\\
&-&{1\over \sqrt 3}( \phi^8)^3
\biggr]~.
\end{eqnarray*}

The critical points in the absence of mass deformations can be taken without loss of generality to be diagonal matrices and hence it is sufficient to look for critical points with nonzero $\phi^8,\phi^3$, and with all the other $\phi$'s vanishing.
The relevant critical point equations are thus

\begin{eqnarray*}\label{cpeqs}
&(\phi^3)^2-(\phi^8)^2=0~,\\
&\phi^8\phi^3=0~.
\end{eqnarray*}
The only solution is thus the trivial solution $\phi^a=0$ and in particular, there is no moduli space of vacua in the undeformed theory. 
Now we deform the superpotential by linear terms for $\phi$. Again, without loss of generality we can add a linear term for just $\phi^3$ and $\phi^8$, so we now consider the superpotential 
$$W={1\over 6} d_{abc}\phi^a\phi^b\phi^c+m_3\phi^3+m_8\phi^8~.$$
We again look for solutions where only $\phi^8,\phi^3$ are activated and the equations for the critical points are modified to
\begin{eqnarray*}\label{cpeqs}
&(\phi^3)^2-(\phi^8)^2+2\sqrt 3 m_8=0~,\\
&\phi^8\phi^3+\sqrt 3 m_3=0~.
\end{eqnarray*}

Clearly, there are two solutions (unless $m_8=m_3=0$, in which case, as we explained above, there is only one solution). The two solutions are related by $\phi^8,\phi^3\to -\phi^8,-\phi^3$, which is nothing but time reversal symmetry 
(there is a choice of time  reversal symmetry that commutes with $SU(3)$, for which all the $\phi^a$ are pseudo-scalars -- this is consistent with~\eqref{pseudos}).
For some special choices of $m_3,m_8$ the assumption that only $\phi^8,\phi^3$ are activated is not justified. Indeed, there are three lines in the $m_3,m_8$ plane where the mass perturbation preserves $SU(2)\times U(1)$ symmetry, namely 

\begin{itemize}
\item $m_3=0$. In this case we see that the solution for $m_8>0$ has nonzero $\phi^8\sim\pm\sqrt m_8  $ and $\phi^3=0$. For $m_8<0$, $\phi^3$ is nonzero and $\phi^8=0$. Therefore the global $SU(2)\times U(1)$ symmetry is
broken to $U(1)\times U(1)$ for $m_8<0$ and unbroken for $m_8>0$. 
\item $m_3=-\sqrt 3 m_8$. Repeating the analysis, one find that the global $SU(2)\times U(1)$ symmetry is
broken to $U(1)\times U(1)$ on the half-line with $m_8>0$.  
\item $m_3=\sqrt 3 m_8$.  One again finds that the global $SU(2)\times U(1)$ symmetry is
broken to $U(1)\times U(1)$ on the half-line with $m_8>0$.  
\end{itemize}  

When the symmetry $SU(2)\times U(1)$ is spontaneously broken to $U(1)\times U(1)$ there is a $\C P^1$ sigma
model at low energies. The $\C P^1$ is parameterized by additional scalars that are massless. Time reversal symmetry acts as an antipodal map on the target space.

\section{Further Checks of the $\mathcal{N}=2$ Dualities}\label{appendix:duality}
We want to analyze in detail the duality we stated in section 3
\begin{equation}
U(1)_0+{\rm charge \ 2} \longleftrightarrow U(1)_2\otimes\left[U(1)_{3/2}+{\rm charge\ 1}\right]~ \label{eq:dualitysmall}
\end{equation}
 In particular, we can try to compare the superconformal index, which for a 3d SCFT with flavor symmetry $U(1)^N$, is defined by~\cite{Imamura:2011su,Kapustin:2011jm}
\begin{equation}
I_{\mathcal{T}}(m;q,\zeta) = \Tr_{\mathcal{H}} \left((-1)^F e^{-\beta H}q^{(E+j_3)/2} \prod_{a} \zeta_a^{e_a}\right)
\end{equation}
where $E$, $j_3$, $e_a$ are energy, the third component of the angular momentum rotating $S^2$, and flavor charges. $\zeta_a$ is the fugacity corresponding to each global symmetry, with $a$ running from $1$ to $N$. Trace is taken over the Hilbert space $\mathcal{H}$ on $S^2$ at a certain magnetic flux $m$ background. Since the generator for $E+j_3$ commutes with the Hamiltonian given by
\begin{equation}
H = \{Q,Q^\dagger\} = E -R -j_3
\end{equation}
($R$ is the R charge of the state) thus we have a fugacity $q$ corresponding to this symmetry. However, instead of working in the fugacity basis and manipulating hyper-geometric functions, it is much easier to work in the charge basis. We refer the reader to \cite{Dimofte:2011py} for more details. Here we just mention a few useful points related to our discussion. We can rewrite the index of the theory $\mathcal{T}$, $I_{\mathcal{T}}(m;q,\zeta) = \sum_{e\in \mathbb{Z^N}} I_{\mathcal{T}}(m,e;q)\zeta^e$, and work with the index at a fixed background magnetic flux $m$ and electric charge $e$. There is an $Sp(2N,\mathbb{Z})$ action on the 3d SCFT with flavor symmetry $U(1)^N$. Accordingly, index transforms as
\begin{equation}
I_{g\circ\mathcal{T}}(g\gamma;q) = I_{\mathcal{T}}(\gamma;q)
\end{equation}
for $g\in Sp(2N,\mathbb{Z})$ and $\gamma = (m,e)^T$ is the symplectic charge vector.

A theory $T_M$ for any 3 manifold M can be built from the basic block $T_\Delta$, the theory for a single ideal tetrahedron~\cite{Dimofte:2011py} consisting only of a single free chiral multiplet. There is a triality enjoyed by the tetrahedron index
\begin{equation}
I_{\Delta}(m,e;q) = (-q^{1/2})^{-e} I_{\Delta} (e,-e-m;q) = (-q^{1/2})^mI_{\Delta}(-e-m,m;q) \label{eq:triality}
\end{equation}
This simply comes from the particle vortex duality
\begin{equation}
I_\Delta(m,e;q) = I_{\sigma_e ST\circ\Delta}(m,e;q) = I_{(\sigma_eST)^2\circ \Delta}(m,e;q)
\end{equation}
Before calculating the index on both sides of the duality, it is important to understand first how our theory on both sides are related to the tetrahedron theory $T_\Delta$. We now show that the duality \eqref{eq:dualitysmall} follows simply from the particle vortex duality $\Delta \leftrightarrow (ST)^2 \Delta$, which we writes as 
\begin{equation}
|D_A\phi|^2 - \frac{1}{2} AdA + \ldots \leftrightarrow |D_b\tilde{\phi}|^2 + \frac12 bdb + 2bdc + cdc +2cdA\ldots
\end{equation}
To avoid clutter, we normalized the Chern-Simons term such that minimal allowed ones looks like $AdA$ and $2AdB$ (i.e. the usual $1/4\pi$ factor is implicit). And $\ldots$ denotes the super-partner parts. Also, we use the upper case and lower case letter for $U(1)$ background and dynamical gauge field respectively. We rename $c\rightarrow c-b$, and integrate out $c$ on the right hand side, we have
\begin{equation}
|D_b\tilde{\phi}|^2 - \frac12 bdb -2bdA  -AdA\ldots
\end{equation}
Now on both sides, we rescale $A \rightarrow 2A$ followed by $ST^2$ operation.
\begin{equation}
|D_{2a}\phi|^2 + 2adA + \ldots \longleftrightarrow |D_b\tilde{\phi}|^2 - \frac12 bdb -4bdc-2cdc+2cdA + \ldots 
\end{equation}
Now if we rename $c\rightarrow c-b$ on the right hand side, we are left with
\begin{equation}
|D_{2a}\phi|^2 + 2adA + \ldots \longleftrightarrow |D_b\tilde{\phi}|^2 + \frac32 bdb -2bdA-2cdc+ 2cdA + \ldots \label{eq:dualityB}
\end{equation}
Therefore, we end up with a single chiral coupled to $U(1)_{3/2}$ tensored with a decoupled $U(1)_{-2} \cong U(1)_{2}$ on the right hand side, which is dual to $U(1)_0$ with a single charge 2 chiral on the left hand side. Note that we can integrate out dynamical field $c$ on the right hand side, and the last two terms just yield $\frac12 AdA$.

Now it is pretty straightforward to calculate the index on both sides of Eq.~\eqref{eq:dualityB}, denoted as $I_{U(1)_{3/2}\otimes U(1)_2} (m,e)$ and $I_{U(1)_{0}}(m,e)$. Knowing the relation between the theory $\mathcal{T}$ and $\Delta$,
\begin{equation}
I_{\mathcal{T}}(m,e) = I_\Delta(m_\Delta,e_\Delta) 
\end{equation}
we just need to find $(m_\Delta,e_\Delta)$ in terms of $(m,e)$. To this end, recall that the presence of Chern Simons terms modifies Gauss' law.\footnote{
	A toy example to see why this happens: consider the Lagrangian $
	\mathcal{L} = \frac{k}{4\pi} \epsilon^{\mu\nu\rho} A_\mu \partial_\nu A_\rho- A_\mu J^\mu$. The variation of an Chern-Simons term is $\frac{k}{4\pi}2\int_{S^1}\delta A \int_{S^2}F $, where $m = \frac{1}{2\pi} \int F \in \mathbb{Z}$ is the magnetic flux. Obviously each magnetic flux carry gauge charge $k$. And the equation of motion is $\frac{k}{4\pi} \epsilon^{\mu\nu\rho}F_{\nu\rho} = J^\mu$ simply tells that magnetic field is proportional to the matter field charge density. So monopole operator has to be dressed with chiral matter operators in order to have gauge charge zero.
	}  A state with nonzero gauge flux $M$ is not gauge invariant. Therefore, we require a state with gauge flux $M$ and flavor flux $m$ dressed with chirals of charge $e_\Delta$ to be gauge invariant, and has flavor charge $e$. The magnetic flux felt by the chiral is denoted as $m_\Delta$. Gauge invariance then gives us the relation between them.

In particular, the theory on the left hand side of \eqref{eq:dualityB} has index 
\begin{equation}
I_{U(1)_{0}}(m,e) = I_\Delta (2e, -e - \frac{m}{2})
\end{equation}
The theory on the right hand side has index
\begin{equation}
I_{U(1)_{3/2}\otimes U(1)_2} (m,e) = I_\Delta(-e + \frac{m}{2},2e)
\end{equation}
which equals to $I_{U(1)_{0}}(m,e)$ up to affine shift of the charge due to the triality property of ideal tetrahedron index \eqref{eq:triality}\cite{Dimofte:2011py}, an indication that the duality indeed follows from the particle vortex duality.

The duality \eqref{eq:dualitysmall} should also be subject to the check of $S^3$ partition function. For 3d $\mathcal{N}=2$ gauge theory on $S^3$, partition function from the localization~\cite{Pufu:2016zxm,Jafferis:2010un} is
\begin{equation}
Z_{S^3}[\Delta_i,\Delta_t] = \frac{1}{|\mathcal{W}|} \int_{\text{Cartan}} d\sigma \prod_{a} \big[e^{i\pi k_a \mathrm{tr}\  (\sigma^a)^2-2\pi \Delta_t\mathrm{tr}\  \sigma^a} det_{Ad}(2\sinh(\pi\sigma^a))\big] \prod_{i}det_{R_i}e^{\mathit{\ell}(1-\Delta_i+i\sigma)},\label{Eq:ZS3}
\end{equation}
where the function $\ell(z)$ is given by 
\[\ell(z) = -z\log(1-e^{2\pi i z})+\frac{i}{2}(\pi z^2+\frac{1}{\pi}\text{Li}_2(e^{2\pi iz})) - \frac{i\pi}{12} \]
where $\sigma_a$ is the adjoint scalar in the vector multiplet of the corresponding $U(1)$ gauge group. $\Delta_i$ is the R charge for $i$-th chiral field. $\Delta_t$ is the topological charge and corresponds to pure imaginary FI parameter. And free energy is defined by $F_{S^3} = -\log|Z_{S^3}|$. It flows to a SCFT at IR fixed point where free energy is maximized with respect to R charge $\Delta_i$ and topological charge $\Delta_t$.

In particular, for the theory $U(1)_0$ with charge 2 chiral on the left hand side of the duality, we have
\begin{equation}
Z_{S^3}[\Delta,\Delta_t] = \int_{-\infty}^{+\infty} d\sigma  \big[e^{-2\pi \Delta_t   \sigma  }\big] e^{\mathit{\ell}(1-\Delta+i2\sigma)},
\end{equation}

The only global symmetry is $U(1)_t$, and only $\Delta_t$ maximization is needed. We obtain numerically,
\begin{equation}
F_{S^3} = 0.989539\ldots,\qquad \Delta_t \to 0
\end{equation}
For the theory $U(1)_2 \otimes [U(1)_{3/2} + \text{charge 1}]$
we have
\begin{equation}
Z_{S^3}[\Delta,\Delta_t] = \int_{-\infty}^{+\infty} d\sigma  \big[e^{i\pi \frac32 \sigma^2 -2\pi \Delta_t   \sigma  }\big] e^{\mathit{\ell}(1-\Delta+i\sigma)} \times \int_{-\infty}^{+\infty} d\sigma  \big[e^{i\pi 2 \sigma^2 -2\pi \Delta_t   \sigma  }\big]
\end{equation}
After minimization, we get
\begin{eqnarray}
F_{S^3} = 0.989539\ldots,\qquad  \Delta \to 1/3, \qquad \Delta_t\to 0
\end{eqnarray}
We see that the $S^3$ partition function matches nicely.

Finally, we examine the duality discussed in the section \ref{sec:sym_enhancement}. It tells us $\mathcal{N}=2$ $U(1)$ gauge theory with two chirals of the charge $+1$ has a $SU(3)$ enhancement of the global symmetry from $SU(2) \times U(1)$. Consequently, the Weyl symmetry group will be $S_3$ instead of $S_2$. We can check this $S_3$ by looking at the index. We denote the electric charge and magnetic charge of the gauge invariant operator to be $e_i$ and $m_i$, $i = 1,2$ corresponding to the maximal torus $U(1)^2$ of $SU(3)$. As before, we need to find $e_{\Delta_1}$, $e_{\Delta_2}$, $m_{\Delta_1}$, $m_{\Delta_2}$ in terms of  $e_i$ and $m_i$, such that
\begin{equation}
I_{\mathcal{T}}(m_1,e_1;m_2,e_2;) = I_{\Delta}(m_{\Delta_1},e_{\Delta_1}) I_{\Delta}(m_{\Delta_2},e_{\Delta_2})
\end{equation}
Again, by requiring the gauge invariance of the states, we obtain
\begin{equation}
I_{\mathcal{T}}(m_1,e_1;m_2,e_2;) = I_\Delta(e_1 - m_2, \frac12(-e_1 - e_2 - m_1 + m_2)) I_{\Delta}(e_1 + m_2, 
\frac12(-e_1 + e_2 - m_1 - m_2))\label{eq:indexCP1}
\end{equation}

We now show explicitly $I_{\mathcal{T}}(m_1,e_1;m_2,e_2;)$ is invariant under the Weyl group action. One example of $S^3$ action on the charge $m_i$ and $e_i$ is given by
\begin{align}
&e'_{1} = \frac12 (-e_1 - e_2) 
&e'_2 =\frac12 (-3 e_1 + e_2)\\
&m'_{1} = \frac12 (-m_1 - 3 m_2)
&m'_2 =\frac12 (-m_1 + m_2)
\end{align}
Note that we use $2T_3$ and $2T_8/\sqrt{3}$ to generate the Cartan $U(1)^2$. So there is no $\sqrt{3}$ in the transformation above.

\begin{equation}
I_{\mathcal{T}}(m'_1,e'_1;m'_2,e'_2;) = I_\Delta(\frac12(-e_1 - e_2 + m_1 - m_2), e_1 + m_2) I_{\Delta}(
\frac12 (-e_1 - e_2 - m_1 + m_2), \frac12 (-e_1 + e_2 + m_1 + m_2))
\end{equation}
which is equal to \eqref{eq:indexCP1} due to the triality property of the ideal tetrahedron index. Other elements of the Weyl group can be checked in the same way.
\bibliographystyle{JHEP}
\bibliography{references}
\end{document}